\documentclass{jfm}
\usepackage{graphicx}
\usepackage{epstopdf, epsfig}
\usepackage{caption,subcaption}
\usepackage{color}

\newcommand{\tn}[1]{{\color{black}#1}}

\shorttitle{Rare transitions to thin-layer turbulent condensates }
\shortauthor{A. van Kan, T. Nemoto and A. Alexakis}

\title{Rare transitions to thin-layer turbulent condensates}

\author{Adrian van Kan \aff{1}
  \corresp{\email{avankan@lps.ens.fr}},
  Takahiro Nemoto \aff{2}
 \and Alexandros Alexakis\aff{1}}

\affiliation{\aff{1} Laboratoire de Physique de l'Ecole normale sup\'erieure, ENS,
Universit\'e PSL, CNRS, Sorbonne Universit\'e, Universit\'e Paris-Diderot,
Sorbonne Paris Cit\'e, Paris, France
\aff{2} Philippe Meyer Institute for Theoretical Physics, Physics Department, \'Ecole Normale Sup\'erieure \& PSL Research University, 24 rue Lhomond, 75231 Paris Cedex 05, France
}

\begin{document}

\maketitle

\begin{abstract}
Turbulent flows in a thin layer can develop an inverse energy cascade leading to spectral condensation of energy when the layer height is smaller than a certain threshold. These spectral condensates take the form of large-scale vortices in physical space.
Recently, evidence for bistability was found in this system close to the critical height: 
depending on the initial conditions, the flow is either in a condensate state with most of the energy in the two-dimensional (2-D) large-scale modes,
or 
in a \tn{three-dimensional (3-D) flow state} with most of the energy in the small-scale modes. 
This bistable regime is characterised by the statistical properties of random and rare transitions between these two locally stable states.
Here, we examine these statistical properties 
in thin-layer turbulent flows, where the energy is injected by \tn{either} stochastic \tn{or} deterministic forcing.
To this end, by using a large number of direct numerical simulations (DNS), we measure 
the decay time $\tau_d$ of the 2-D condensate to 
\tn{3-D flow state} and the build-up time $\tau_b$ of the 2-D condensate.
We show that both of these times $\tau_d,\tau_b$ %
follow an exponential distribution 
with mean values increasing faster than exponentially as the layer height approaches the threshold. 
We further show that the dynamics of large-scale kinetic energy may be modeled by a stochastic Langevin equation. 
From time-series analysis of DNS data, we determine the effective potential that shows two minima corresponding to the 2-D and 3-D states when the layer height is close to the threshold.
\end{abstract}  


\section{Introduction}\label{sec:intro}
Turbulence is ubiquitous in the universe, from stars to tea cups. Many astrophysical and geophysical turbulent flows, such as planetary oceans and atmospheres, are subject to geometrical constraints, e.g. thinness in one spatial direction \citep{pedlosky2013geophysical}. Such constraints significantly change the properties of the flow, which therefore deviate from those of classical three-dimensional (3-D) homogeneous and isotropic turbulence. Fully 3-D turbulence is characterised by a forward cascade of energy from large to small scales \citep{frisch1995turbulence}, while in two dimensions (2D), 
an inverse energy cascade from small to large scales occurs due to additional inviscid invariants such as enstrophy 
\citep{bofetta2012twodimensionalturbulence}. 
Turbulence in thin layers combines properties of both cases,
as large-scale dynamics are constrained to be 2-D, while small-scale dynamics are not \tn{\citep{smith1996crossover, celani2010morethantwo, benavides_alexakis_2017, musacchio2017split}}. As a consequence, in thin-layer turbulence, energy may cascade both to small and large scales depending on the layer height $H$: \tn{Above a critical height $H_{_{3D}}$ there is no inverse cascade, while below this critical height an inverse cascade develops, whose amplitude (measured by the inverse energy flux) increases continuously. Furthermore, a second critical height $H_{_{2D}} < H_{_{3D}}$ was 
observed where the flow become exactly two-dimensional and no forward cascade was observed.}
Similar transitions towards an inverse cascade occur in rotating turbulence \tn{\citep{smith1999transfer,deusebio2014dimensional}},
stratified turbulence \tn{\citep{metais1996inverse,marino2013inverse,marino2014large,sozza2015dimensional}} and magnetohydrodynamic systems \tn{\citep{alexakis2011two, seshasayanan2016critical,seshasayanan2014edge}}, among others. 
(see the review articles \tn{\citep{alexakis2018cascades, pouquet2018helicity}}).

In a finite domain, an inverse cascade saturates at late times, 
forming a condensate in which the energy is concentrated in the largest scales.  This condensation has been extensively studied in 2-D turbulence \citep[see][]{hossain1983long, smith1993bose, smithr1994finite, chertkov2007dynamics}.
In quasi-2D systems it has been observed in
rapidly rotating convection \citep{rubio2014upscale,favier2019subcritical}, rotating turbulence \citep{ alexakis2015rotating,yokoyama2017hysteretic,seshasayanan2018condensates} and thin-layer turbulence \citep{xia2011upscale,vankan2018condensates}. 
In many of these cases, the amplitude of the condensate state (measured by the energy in the large scales) has been shown to
vary discontinuously with the system parameters.
Furthermore, close to the transition, bistability has been observed: the system was either attracted or not to the condensate state depending on the initial conditions.
In particular in thin-layer flows, for values of $H$ close to $H_{_{3D}}$, the system was attracted to either a 2-D condensate state (where most of the energy is concentrated in two counter-rotating, large-scale, 2-D vortices) or a \tn{3-D flow state} (where energy is mostly contained in 3-D small-scale fluctuations)  \tn{\citep{vankan2018condensates, musacchio2019condensate}}. 
The bistability in this system was accompanied by sudden `{\it jumps}' between these two states. These transitions occur randomly with the waiting times that  are, presumably, stochastic, following a statistical distribution that characterises the bistable regime.  

In this paper, we present the first analysis of the statistical properties of thin-layer turbulence close to the critical height. We use a very large number of direct numerical simulations (DNS) and calculate the probability distribution functions (PDFs) of the transition times: the {\it decay time} $\tau_d$ from a 2-D condensate state to a \tn{3-D flow state} and the {\it build-up time} $\tau_b$ from a \tn{3-D flow state} to a 2-D condensate state. We examine their dependence to $H$ and attempt to model the transitions in terms of a particle in a one-dimensional  potential using a Langevin equation.

\section{Setup and results from direct numerical simulations} \label{sec:dns}

In this study, we consider forced incompressible 3-D flow in a triply periodic domain of dimensions $L\times L\times H$ with $H\ll L$. The setup is identical with the one studied in \citep{vankan2018condensates}. The thin direction is referred to as the \textit{vertical} `$z$' direction and the remaining two as the \textit{horizontal} `$x,y$' directions. The flow obeys the incompressible Navier-Stokes equation
\begin{subeqnarray}
\partial_t \mathbf{v} + \mathbf{v}\cdot \nabla \mathbf{v} =& -& \nabla P + \nu \nabla^2 \mathbf{v} + \mathbf{f} \label{eq:Navier_Stokes},\\
\nabla \cdot \mathbf{v} =& 0&,
\end{subeqnarray}
where $\mathbf{v}(\mathbf{x},t)$ is the velocity field, $P(\mathbf{x},t)$ is physical pressure divided by constant density,  $\nu$ is kinematic viscosity and $\mathbf{f}$ is the external body force injecting energy into the flow. In this work, we use two different forcing functions: stochastic $\mathbf{f}_s$ and deterministic $\mathbf{f}_d$. Both forcing functions depend only on $x$ and $y$ and have only $x$ and $y$ components, {\it i.e.,} are two-dimensional-two-component (2D2C) fields. 
In both cases, the force is divergence-free and only acts on a shell of wavenumbers $|\mathbf{k}| = k_f=2\pi/\ell$. The stochastic force is delta-correlated in time, which leads to a fixed mean injection rate $\langle \mathbf{v \cdot \mathbf{f}_s\rangle = }\epsilon$.
\tn{
The deterministic force $\mathbf{f}_d$ is written in terms of the Fourier transform of the velocity field, $\mathbf{\hat{v}}(\mathbf{k})$, as
\begin{equation}
\mathbf{f}_d(\mathbf{x},t) = \epsilon 
\sum_{ \mathbf{k} \in K_F } \frac{\mathbf{\hat{v} }_{2D}(\mathbf{k}) e^{i\mathbf{k}\cdot \mathbf{x}}}{\sum_{ \mathbf{k'} \in K_F } |\mathbf{\hat{v}}_{2D}(\mathbf{k}')|^2} + i (\epsilon k_f^2)^{1/3}\sum_{ \mathbf{k'} \in K_F } \Omega_{\mathbf{k}} \mathbf{\hat{v}}_{2D} (\mathbf{k})e^{i\mathbf{k}\cdot \mathbf{x}},
\label{eq:detforc}
\end{equation} 
where the sums are taken over all modes in the set $K_F$ that consists of all wavenumbers 
with $k_z=0$ and $ k_f \le |{\bf k}| < k_f+1$ and only in-plane components $\hat{\mathbf{v}}_{2D} =(\hat{v}_x,\hat{v}_y,0)$ are forced. There are two terms in this forcing. The first term is responsible 
for injecting energy at a fixed rate $\epsilon$, matching the mean injection rate of $\mathbf{f}_s$. The second term is not injecting energy,
but is responsible for de-correlating the phases of the forced modes at a timescale given 
by $(\epsilon k_f^2)^{-1/3}$. This is achieved by the linear term $i (\epsilon k_f^2)^{1/3}\Omega_{\mathbf{k}} \mathbf{\hat{v}(k)}$ where $\Omega_{\mathbf{k}}$ are time-independent random numbers that are uniformly distributed over $[-1,1]$. They are kept fixed throughout the simulation time and are the same for all simulations.
Since $\Omega_{\mathbf{k}}$ are time-independent, the forcing is indeed deterministic, i.e. is fully determined by the velocity field at any time. Without the second term, 
the flow at late times is attracted to a degenerate state where the velocity and the forcing are strongly correlated so that two strong vortices with opposite signs, of the size of the forcing scale, are formed. 
%
This forcing (\ref{eq:detforc}) has previously been used successfully by \citep{benavides_alexakis_2017}.}

For both forcing functions, the system is characterised by three non-dimensional parameters: the injection scale Reynolds number $\Rey = (\epsilon \ell^4)^{1/3}/\nu$, the ratio between forcing scale and box height $Q=\ell/H$, and the ratio between forcing scale and the
horizontal domain size $K = \ell/L$.
In all simulations, we focus on the horizontal large-scale kinetic energy, defined as 
\begin{equation}
V_{ls}^2 = \sum_{\mathbf{k}, k_z=0 \atop |\mathbf{k}|<k_{max} }
\left[|\hat{v}_x({\bf k})|^2 +  |\hat{v}_y({\bf k})|^2 \right],
\end{equation} 
where $k_{max} = \sqrt{2} \frac{2\pi}{L}$. The simulations performed for this work use an adapted version of the Geophysical High-Order Suite for Turbulence (GHOST) which uses pseudo-spectral methods including 2/3 de-aliasing to solve for the flow in the triply periodic domain, \citep[see][]{mininni2011hybrid}. For all experiments, we fix $K=1/8$ and $\Rey = 203$ at a resolution of $128 \times 128 \times 16$, varying $Q$ over the interval $[\textcolor{blue}{1.4}, 2.0]$. We choose this low resolution and $\Rey$ because very long-duration runs are needed for this study. In addition, we also made runs at a resolution $256\times 256\times 16$ at $\Rey=406$, which qualitatively showed the same dynamics, even though reliable statistical analysis was not done in this case due to longer CPU times required to integrate (\ref{eq:Navier_Stokes}).

\begin{figure}
\centering
\begin{subfigure}[b]{0.33\linewidth}
\includegraphics[width=1\textwidth]{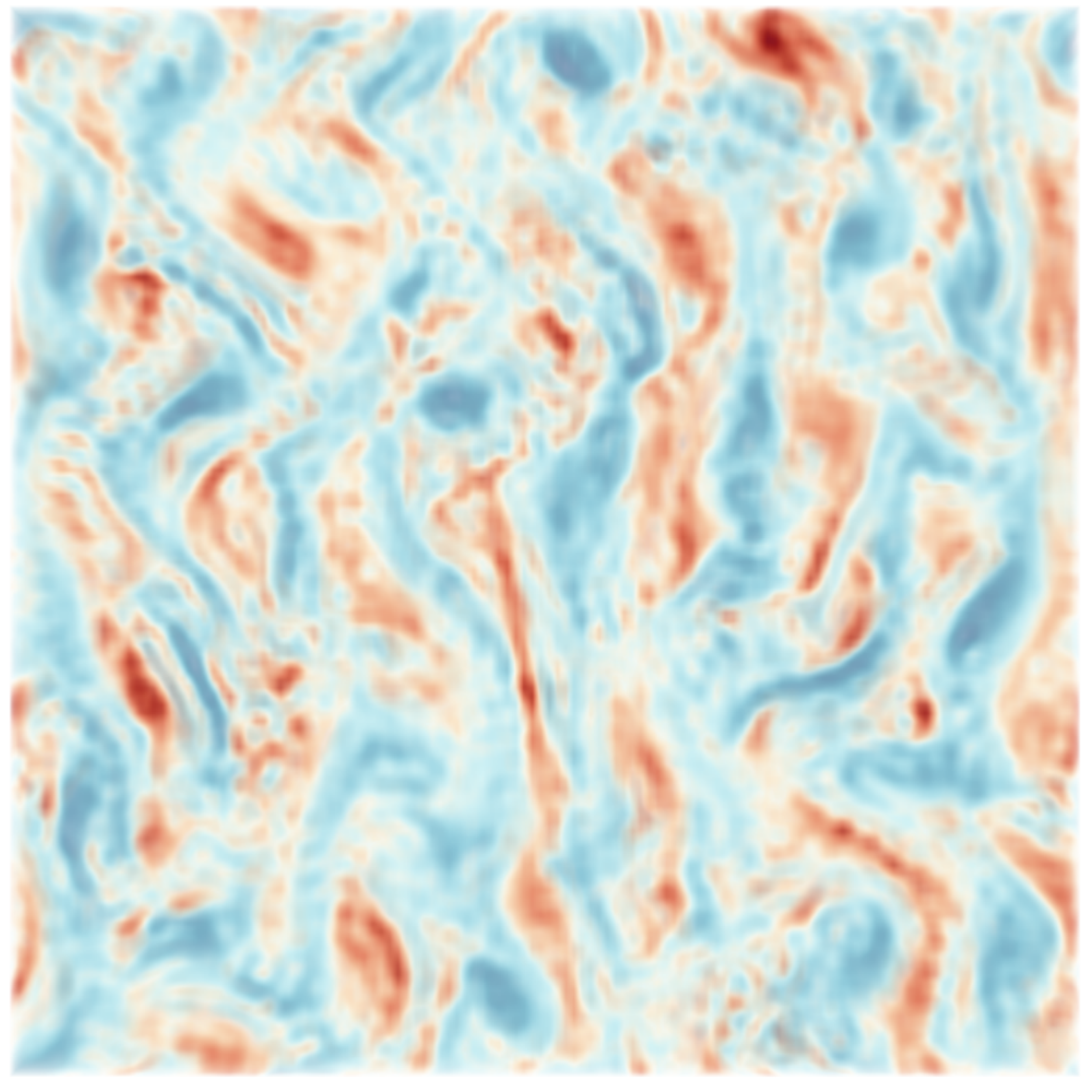}
\caption{3-D flow state}
\end{subfigure}
\hspace{1cm}
\begin{subfigure}[b]{0.33\linewidth} 
\includegraphics[width=1\textwidth]{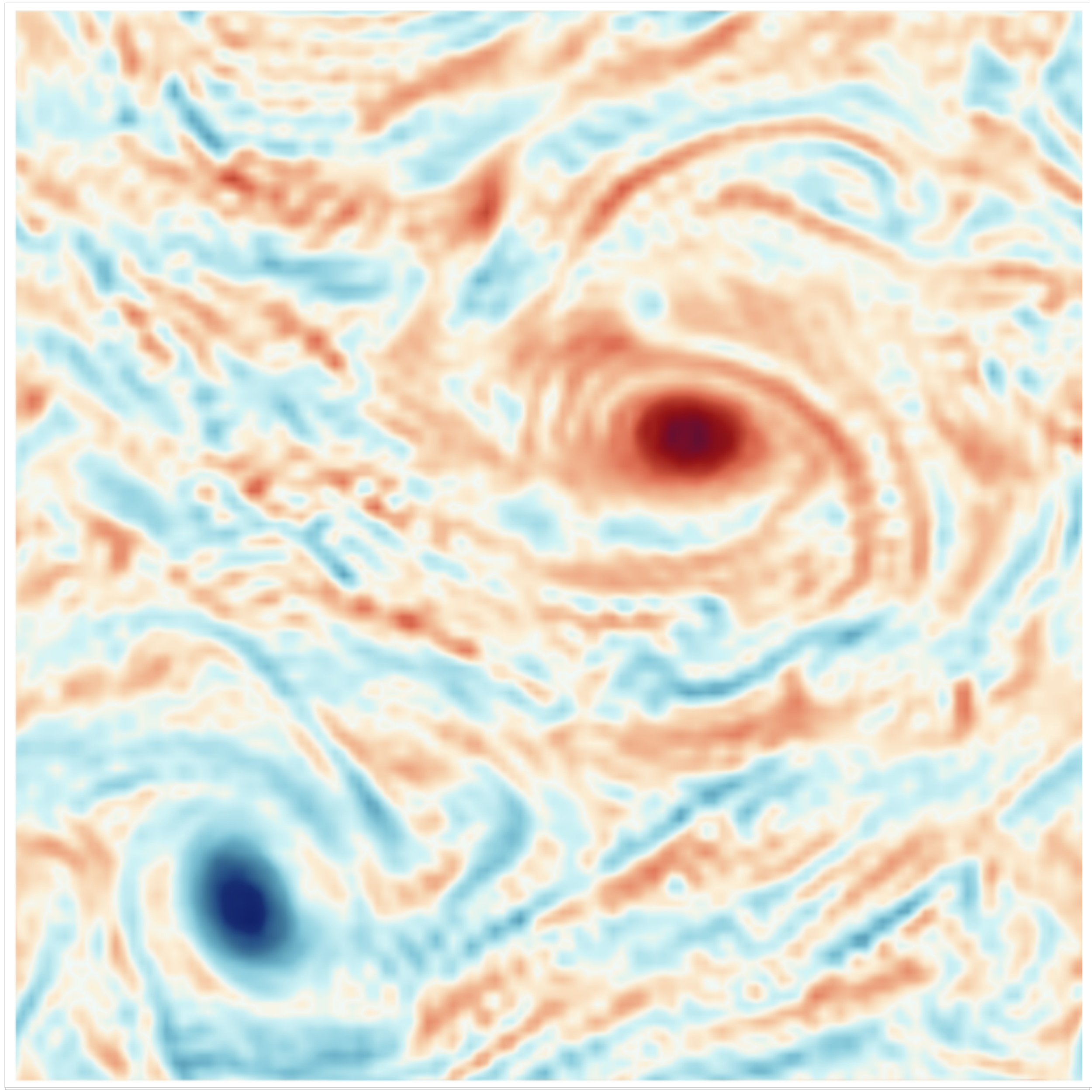}
\caption{2-D condensate state}
\end{subfigure}
\caption{\tn{Top view of the flow field visualised in terms of vorticity (positive in red, negative in blue) in 3-D flow state (left) and 2-D condensate state (right). Both snapshots are taken for stochastic forcing at $Q=1.56$, approximately the middle of the hysteresis loop.}  }
\label{fig:visualisation_vorticity}
\end{figure}

\begin{figure}
\centering
\includegraphics[width=0.38\textwidth]{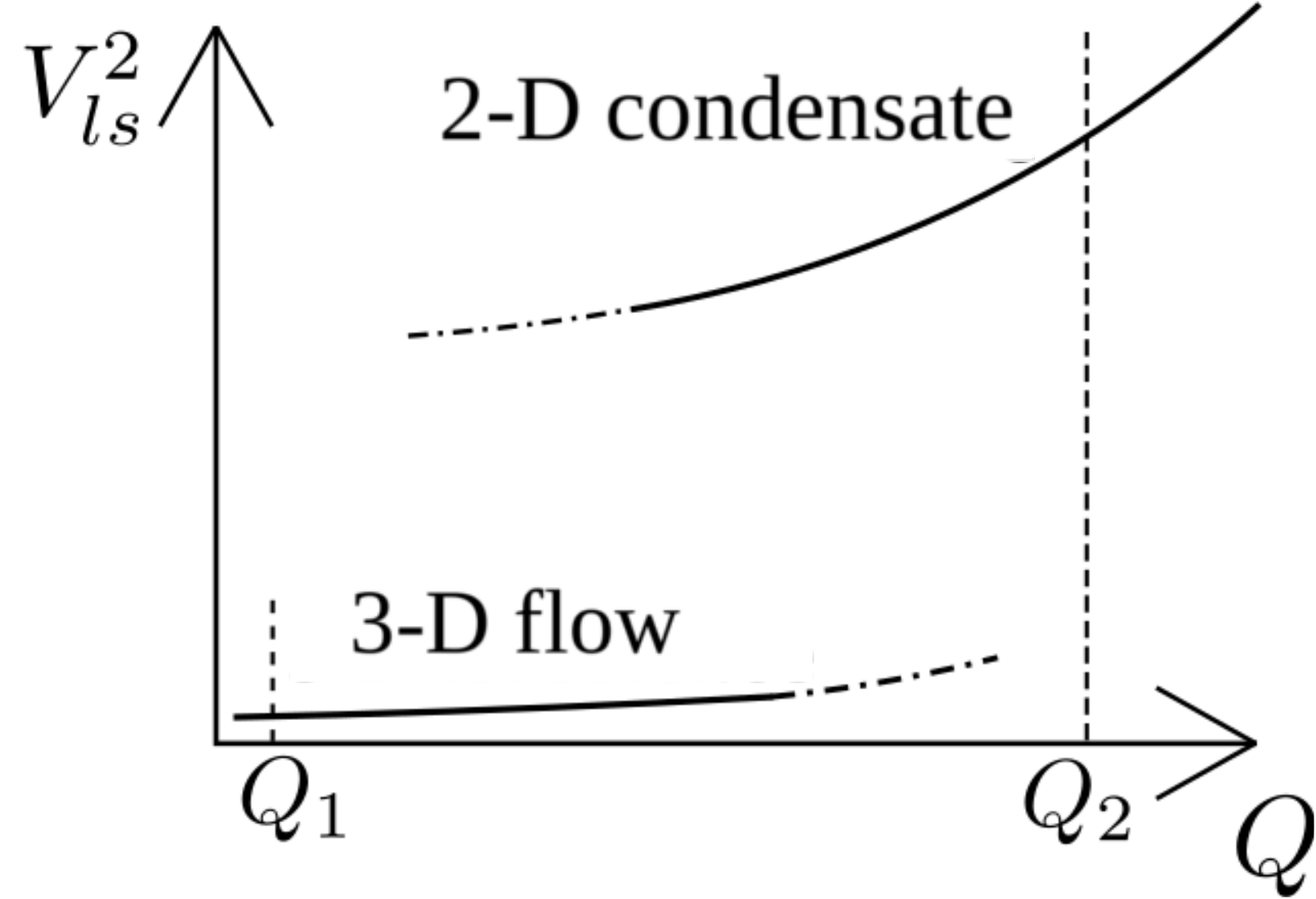}
\caption{Sketch illustrating the hysteresis loop close to the critical height.
Solid and dash-dotted lines represent the stable and unstable branches. We select initial conditions for {\it decay} and {\it build-up} experiments from the typical configurations at sufficiently small and large values of $Q$ (here denoted by $Q_1$ and $Q_2$), respectively.
}
\label{fig:sketch_hyst}
\end{figure}
\begin{figure}
\centering
\begin{subfigure}[b]{0.48\linewidth}
\includegraphics[width=1\textwidth]{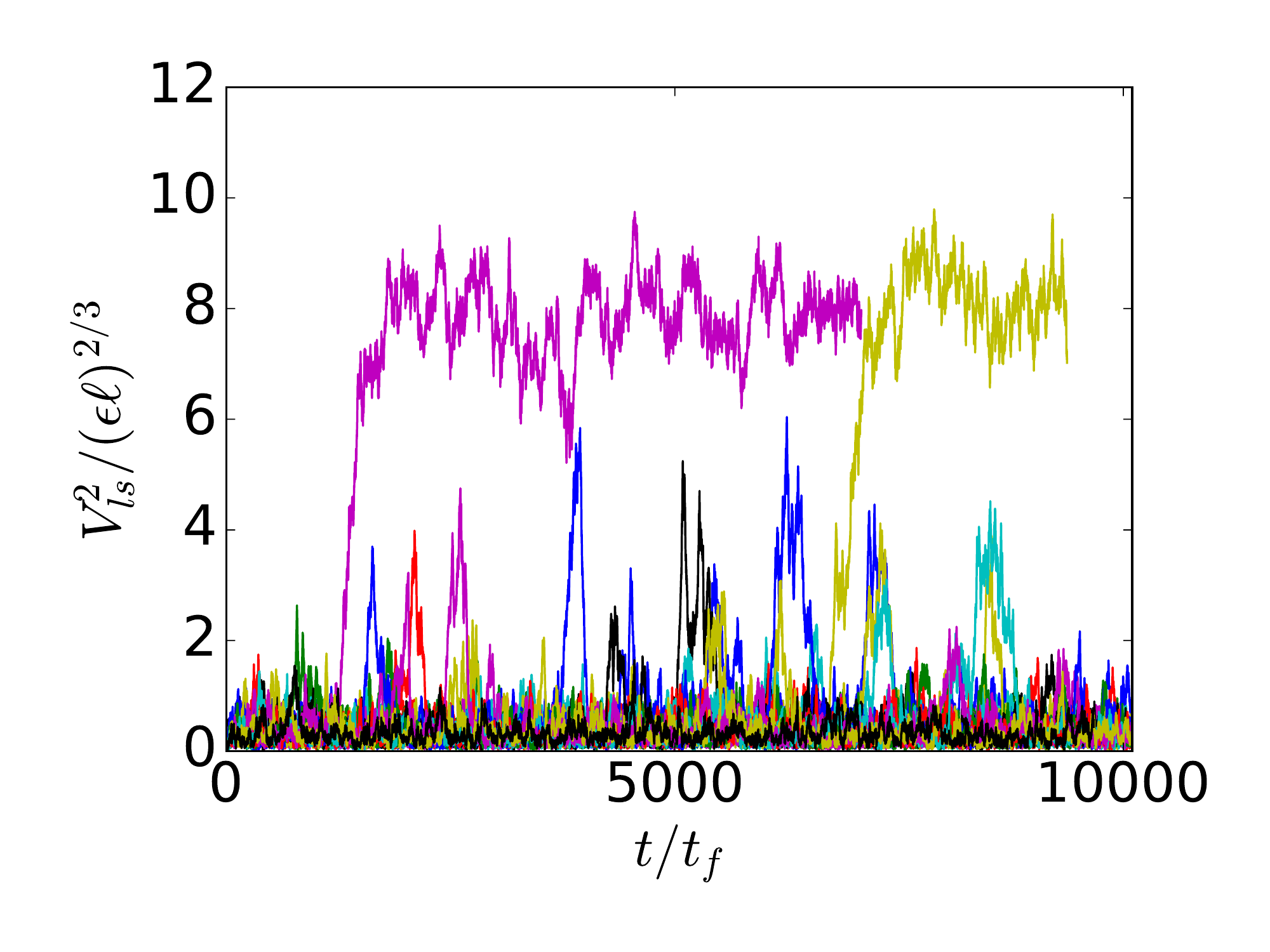}
\end{subfigure}
\begin{subfigure}[b]{0.48\linewidth}                                   
\includegraphics[width=1\textwidth]{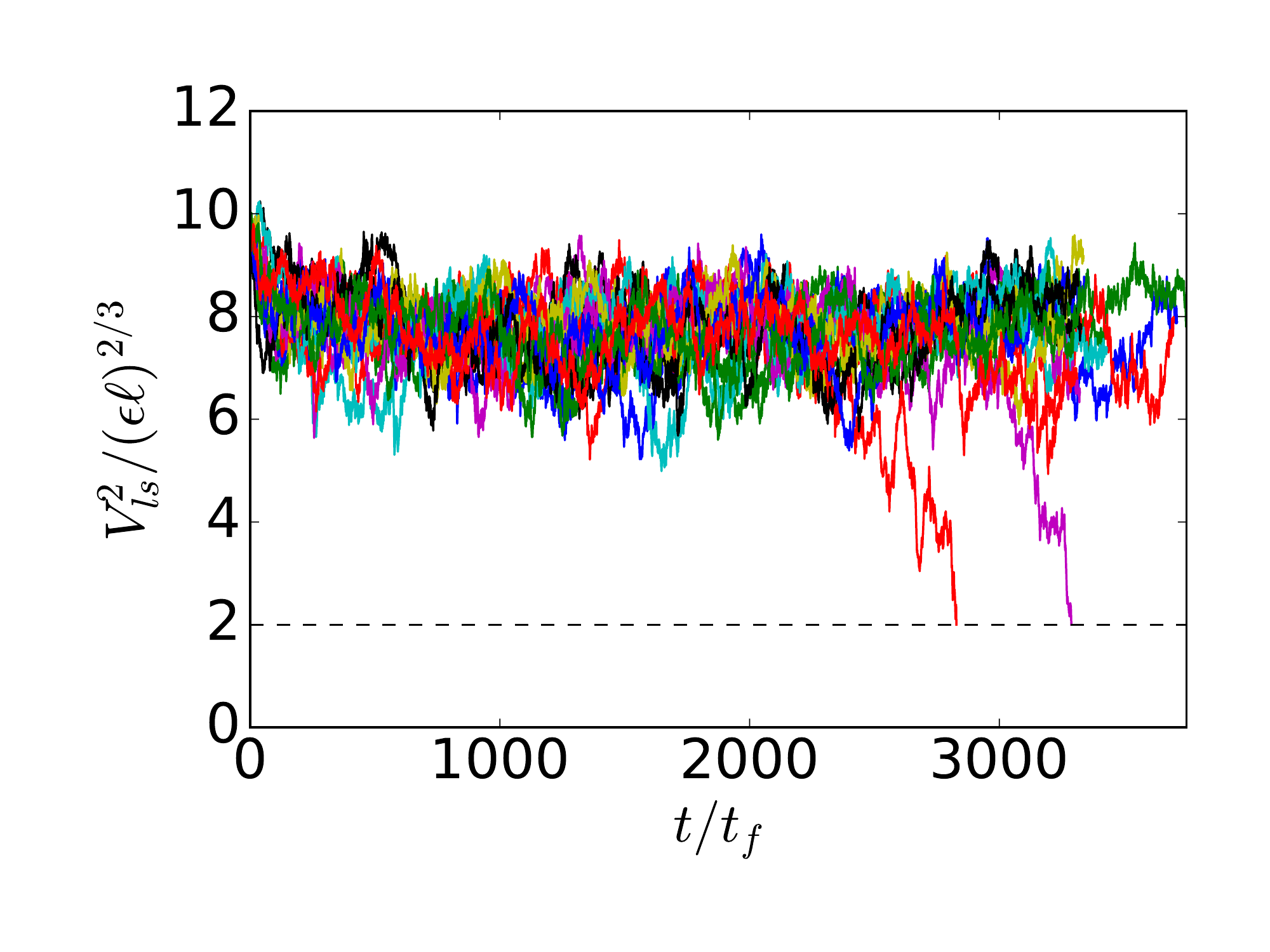}
\end{subfigure}
\caption{Typical time-series of $V_{ls}^2$ for build-up experiments ($Q=1.55$, left) 
and decay ($Q=1.556$, right)
experiments ($\mathbf{f}=\mathbf{f}_s$). 
 We define a build-up time $\tau_b$ (or a decay time $\tau_d$) as the time when $V_{ls}^2$ grows (or drops) to its condensate mean value (or a small threshold $\approx2(\epsilon \ell)^{2/3}$). These $\tau_b$ and $\tau_d$ fluctuate and their statistics are of our interest. }
\label{fig:typical_timeseries}
\end{figure}

\tn{
Depending on $Q$ and the initial conditions, the system is
attracted either to the 2-D condensate state or the 3-D flow state.
Two snapshots of the vorticity for these two states are displayed in figure \ref{fig:visualisation_vorticity}
where in the left panel the flow is in the 3-D state and has a small value of $V_{ls}$ while in the right panel the flow is in the  2-D condensate state and has large value of $V_{ls}$.
Figure \ref{fig:sketch_hyst} shows a sketch of these states in the $Q-V_{ls}$ plane. The upper branch corresponds to the 2-D condensate state while the lower branch
corresponds to the \tn{3-D flow state}. The dash-dotted lines indicate
regions where one of the two states is unstable and the flow can jump from that state to the other.}
%
As detailed below, we perform {\it decay} and {\it build-up} experiments for each value of $Q$. 
In the build-up experiments,
initial conditions corresponding to the \tn{3-D flow state} are used (see figure \ref{fig:sketch_hyst}). In these simulations, we observe the build-up of 2-D condensates after a certain simulation time, which we denote by $\tau_b$. In the decay experiments, initial conditions corresponding to the 2-D condensate state are used. 
The system is evolved until the 2-D condensate decays, where the decay time is denoted by $\tau_d$. 
When decay or build-up events occur, the integration is interrupted and 
the next independent experiment is initiated. 
For the stochastic forcing, runs are started from a fixed initial condition but with different random number sequences, while for deterministic forcing, the initial conditions are altered by a small random perturbation that is different for every run.
Figure \ref{fig:typical_timeseries} illustrates typical time-series for build-up and decay experiments. 
We note that similar procedures have been used to 
determine a critical Reynolds number for turbulent-laminar transitions in pipe flows \citep[see][]{avila2011onset}.

\section{Build-up and decay times} \label{sec:stat_transition}
\begin{figure}
\centering
\begin{subfigure}[hb]{0.47\linewidth}                                   
\includegraphics[width=1\textwidth]{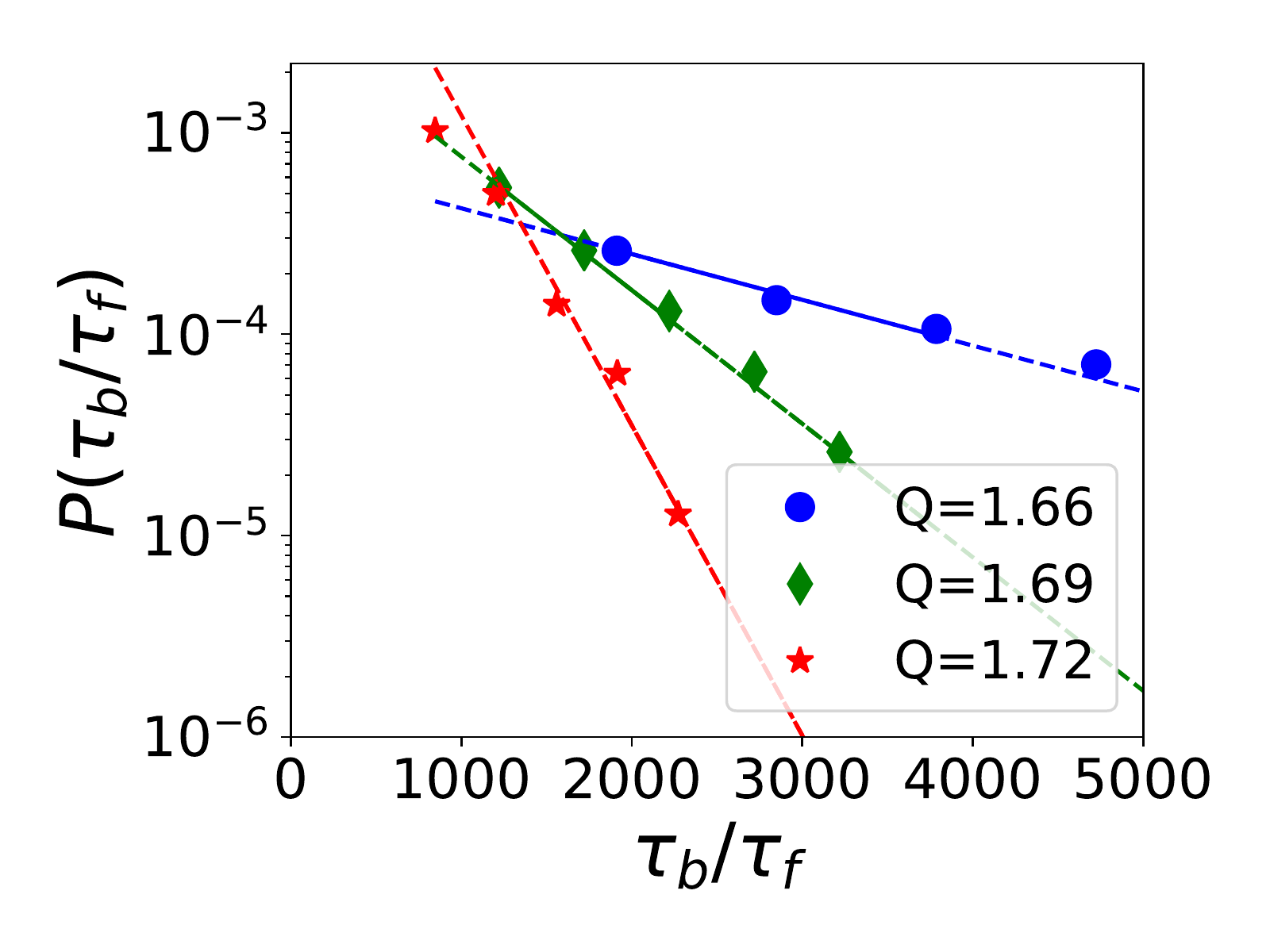} 
\end{subfigure}
\begin{subfigure}[hb]{0.47\linewidth}   
\includegraphics[width=1\textwidth]{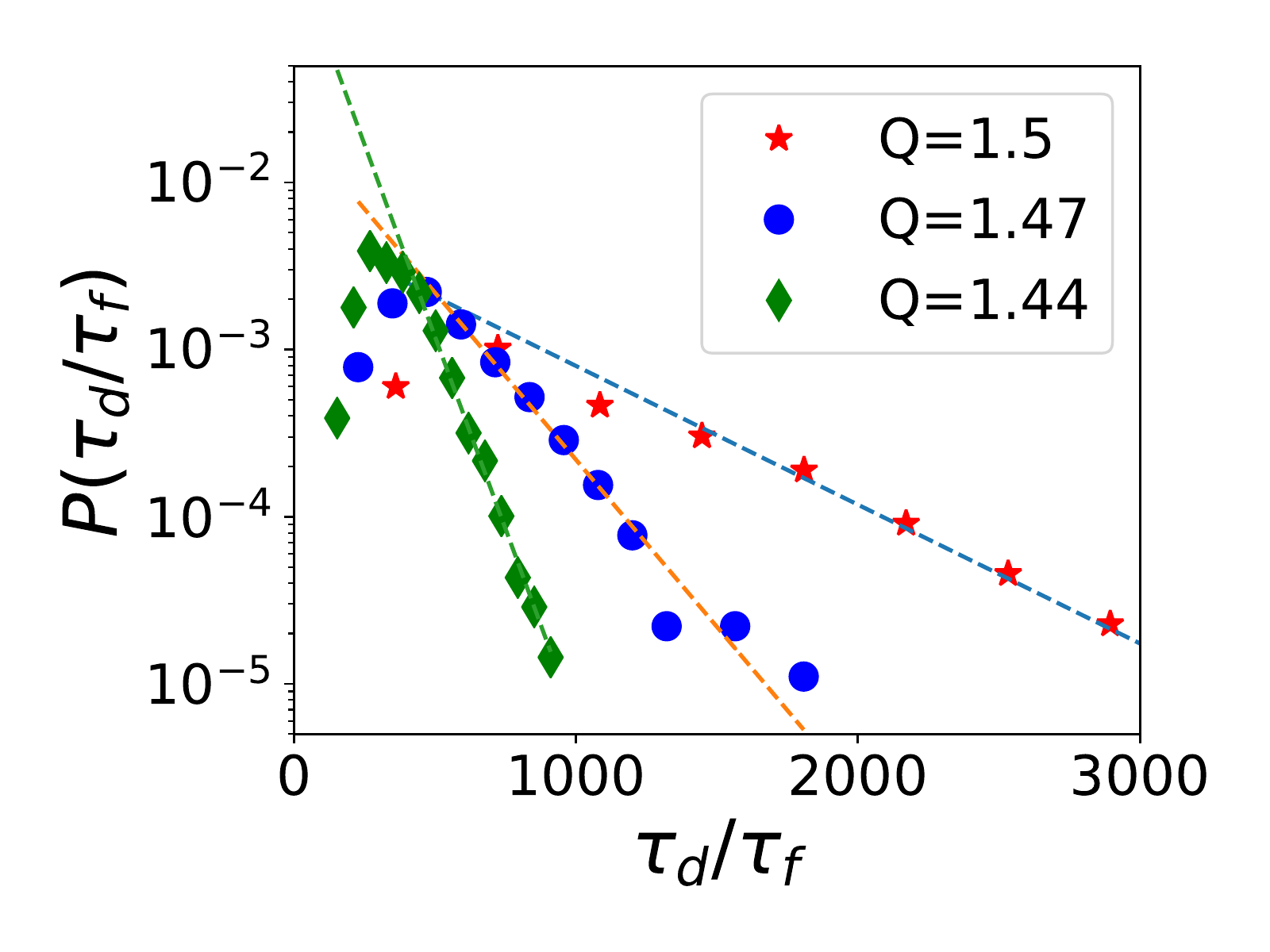} 
\end{subfigure}
\caption{PDFs of the scaled build-up time $\tau_b/\tau_f$ (left) 
and the scaled decay time $\tau_d/\tau_f$ (right)
for different values of $Q$, with $\tau_f=( \ell^2/\epsilon)^{1/3}$. 
The stochastic forcing $\mathbf{f}_s$ is used. The PDFs have exponential tails whose characteristic time scale increases as the transition is approached.}
\label{fig:PDFs}
\end{figure}

\begin{figure}
\centering
\begin{subfigure}[hb]{0.47\linewidth}                          
\includegraphics[width=1\textwidth]{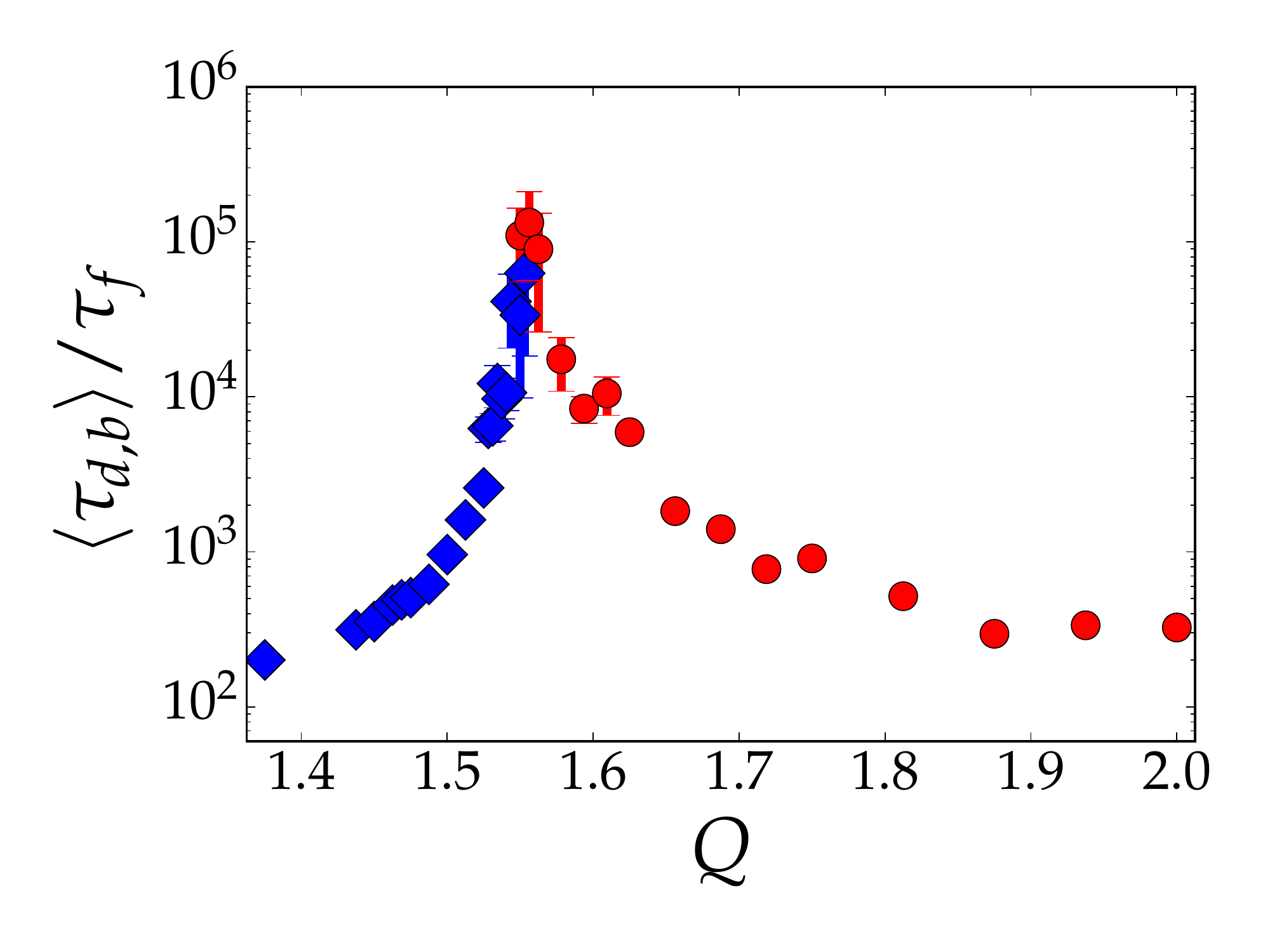} 
\end{subfigure}
\begin{subfigure}[hb]{0.47\linewidth}                                      
\includegraphics[width=1\textwidth]{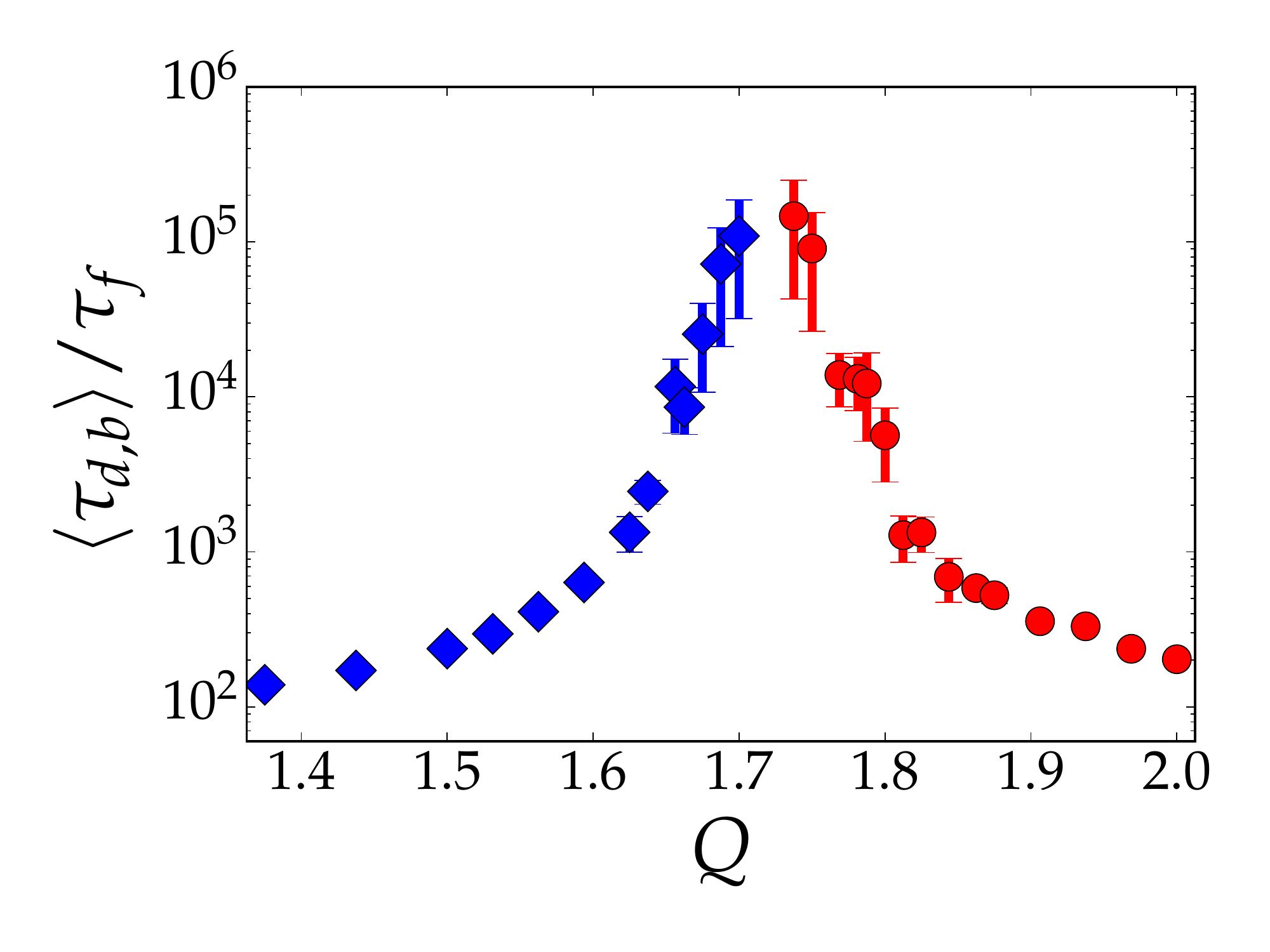} 
\end{subfigure}
\caption{Mean transition times, non-dimensionalised by $\tau_f=( \ell^2/\epsilon)^{1/3}$, for stochastic (left)
and deterministic (right)
forcing as a function of $Q$. Blue diamonds correspond to $\langle \tau_d \rangle$ in the decay experiments and red circles  correspond to $\langle \tau_b \rangle$ in the build-up experiments. Error bars are estimated based on the approximation that the PDF is exactly exponential.}
\label{fig:mean_time}
\end{figure}

We measure the statistical properties of the decay and build-up times $\tau_d,\tau_b$ by \tn{simulating the system for more than ten million ($10^7$) eddy turnover times, using a total of 2.5 million CPU hours. This corresponds to between 40 and over 1000 independent runs per value of $Q$.} 
Figure \ref{fig:PDFs} shows PDFs of $\tau_d$ and $\tau_b$ for the stochastic forcing $\mathbf{f}_s$ for different values of $Q$. All PDFs have an exponential tail, whose slope (in log-linear scale) increases as the transition is approached. The PDFs for the deterministic forcing $\mathbf{f}_d$ 
show qualitatively the same results. \tn{The PDFs remain almost unchanged when half of the data points in the sample are removed, which indicates that they are sufficiently well sampled.} An exponential PDF of waiting times implies that the waiting mechanism can be modeled using a memoryless process, \tn{in which long-time correlations are absent} \citep{billingsley2008probability}. We explore this possibility in Section \ref{sec:markov}.

The resulting mean build-up and decay times, $\langle\tau_b\rangle,\langle\tau_d\rangle$, are shown in figure \ref{fig:mean_time} for stochastic (left panel) and deterministic (right panel) forcing in log-linear coordinates. \tn{The uncertainty in the transition times due to the finite sample size is estimated based on exponential PDFs and shown by the error bars.} The ascending branch (left, in blue) represents the mean decay time $\langle\tau_d\rangle$  and the descending branch (right, in red) represents the build-up time $\langle\tau_b\rangle$. For each case, both transition times increase drastically when a certain value of $Q$ is approached. 
This increase is faster than exponential (super-exponential) and could be either diverging at some critical values $Q_b$ and $Q_d$ or staying finite (similar to what is observed for the decay and split time of turbulent puffs in pipe flows \citep{hof2006finite,hof2008repeller,avila2011onset}). 
%
\tn{
In the former case, where the two times diverge,  
      for $Q< Q_b$ the 2-D condensate will eventually  decay to the 3-D flow state, 
while for $Q>Q_d$ the 3-D state will transition to a condensate. Thus $Q_d$ corresponds to  the smallest value of $Q$ for which a condensate can be observed at late times, while $Q_b$ is the largest value of $Q$ at which 3D-flow state can be observed. Both $Q_b$ and $Q_d$ are close to the value $Q_{3D}\propto 1/H_{_{3D}}$ that 
gives the onset of the inverse cascade as observed in \citep{benavides_alexakis_2017, vankan2018condensates}. 
We note however that in \citep{benavides_alexakis_2017}, where the presence of a hypo-viscosity and the large box limit was considered, no bistability was observed, while in \citep{vankan2018condensates} due to the limited statistics in that work the distinction between  $Q_b$ and $Q_d$ could not be made.  }
%
%
In the latter case \tn{of transition times staying finite despite the super-exponential increase,} a double-exponential function might be used to fit this super-exponential increase, supported by an argument using extreme value statistics \citep{fisher1928limiting,gumbel1935valeurs,goldenfeld2010extreme,goldenfeld2017turbulence}. This can be justified if we assume that the transition to the condensate state is triggered when the maximum value of the small-scale vorticity exceeds a certain threshold value. \tn{Testing this assumption requires knowledge of the entire flow field at times near rare transition events. 
We believe that it would be important to clarify this question in a future study}. 

\begin{figure}
\centering
\includegraphics[width=1\textwidth]{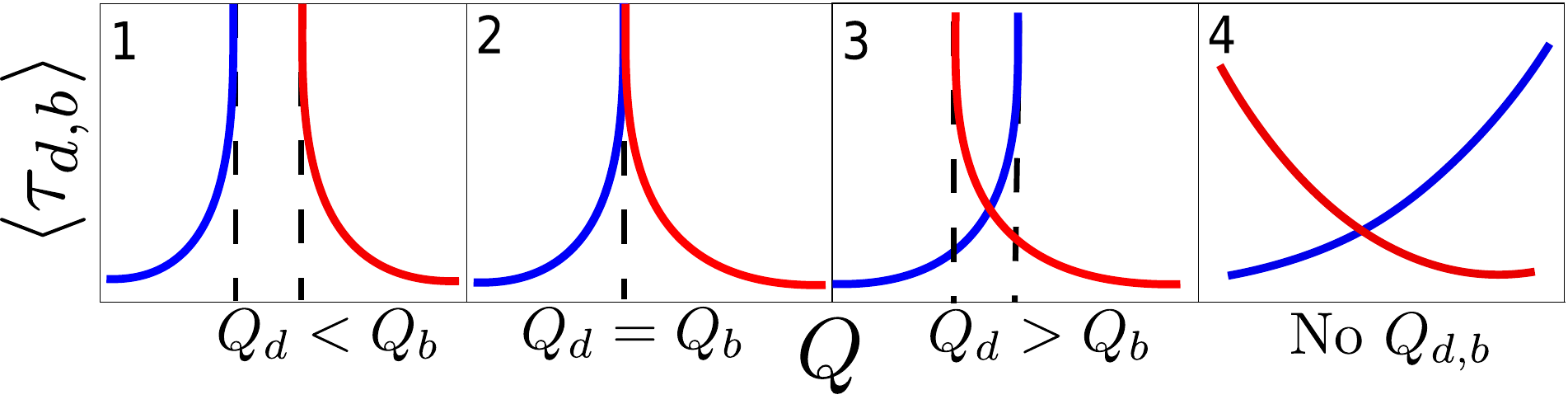}
\caption{Four scenarios for dependence of transition times on $Q$. In each panel, the x-axis represents $Q$ and the y-axis is either the mean decay time $\langle \tau_d\rangle$ (left branch) or the mean build-up time $\langle \tau_b\rangle$ (right branch). Dashed vertical lines indicate $Q_{d,b}$, where the mean transition times diverge. 
}
\label{fig:scenarios}
\end{figure}

Four different scenarios may be envisaged for the $Q$-dependence of the transition times, illustrated in figure \ref{fig:scenarios}. 
In the first scenario, both $\langle \tau_b\rangle$ and $\langle \tau_d\rangle$ diverge at $Q_b, Q_d$ with $Q_d<Q_b$. In the range $(Q_d,Q_b)$, where both transition times diverge, either the \tn{3-D flow} or 2-D condensate state is selected, depending on initial conditions.
In the second scenario, $\langle \tau_b\rangle$ and $\langle \tau_d\rangle$ diverge at the same point, {\it i.e.,} $Q_d=Q_b$. This is an analogue of standard equilibrium phase transitions with a single power-law singularity.
In the third scenario, a crossover is observed, {\it i.e.,} $Q_b<Q_d$, where random transitions between the two states are possible within a finite range of $Q$ between $Q_b$ and $Q_d$.
Finally, in the fourth scenario, a crossover is again observed, but without any divergence of $\langle \tau_b\rangle$ and $\langle \tau_d\rangle$ for finite $Q$. This scenario is compatible with the double-exponential fitting function explained above, where transitions between the two states are possible for all $Q$ (although they are extremely rare). 
From the data in figure \ref{fig:mean_time}, we can see that for the stochastic forcing, the two branches are intersecting around $Q=Q_s\approx1.55$ at a value of around $10^5$ eddy turnover times. Therefore we can exclude cases 1 and 2 for $\mathbf{f}_s$, while all four scenarios are possible for $\mathbf{f}_d$. In addition, it may be proven \citep[see][]{gallet_doering_2015} and has been confirmed numerically \citep{vankan2018condensates}, that beyond a second critical value $Q_{2D}$, the flow two-dimensionalises and the condensate is stable to 3-D perturbations in the long-time limit. Therefore, condensate decay time diverges at least for $Q> Q_{2D}$ and we can exclude a 
double-exponential behavior extending to all $Q$ for $\langle \tau_d\rangle$. Note that based on our \tn{available data}, we cannot exclude that condensate build-up time never becomes infinite. In a future study, rare event algorithms \tn{(see Section \ref{sec:conclusions}) may help elucidating these questions, including the problem of how rare transitions are triggered. }

\tn{An alternative way of reducing the computational cost of the problem could be to consider not the statistics of transition events which become extremely rare, but rather the statistics of bursting events, which can be seen in figure \ref{fig:typical_timeseries} to be significantly more frequent than transition events in the bistable regime. While we have not studied them in very much detail, we have found that waiting times for bursting events also follow exponential PDFs, whose mean increases towards larger $Q$. Bursting events may prove more fruitful than transition events at higher Reynolds numbers and we believe they are a promising object of study for future work. }

\section{Effective Markovian modelling} \label{sec:markov}

%
The exponential PDFs of the transition times (figure \ref{fig:PDFs}) indicate that these times are stochastically determined by a mechanism that \tn{is memoryless, in other words, which leads to vanishing long-time correlations for these times}. Since the transitions are quantitatively characterized by a single macroscopic variable, the horizontal large-scale kinetic energy $V_{ls}^2$ (figure \ref{fig:typical_timeseries}), this observation implies that the dynamics of $V_{ls}^2$ could be described by a Markov (memory-less) process, such as an inertia-less particle moving in a double-well potential in the presence of white noise. Within this effective description, the transitions are characterized as rare jumps of the particle between the two wells of the potential.

\begin{figure}
\centering
\begin{subfigure}[hb]{0.32\linewidth}
\includegraphics[width=1\textwidth]{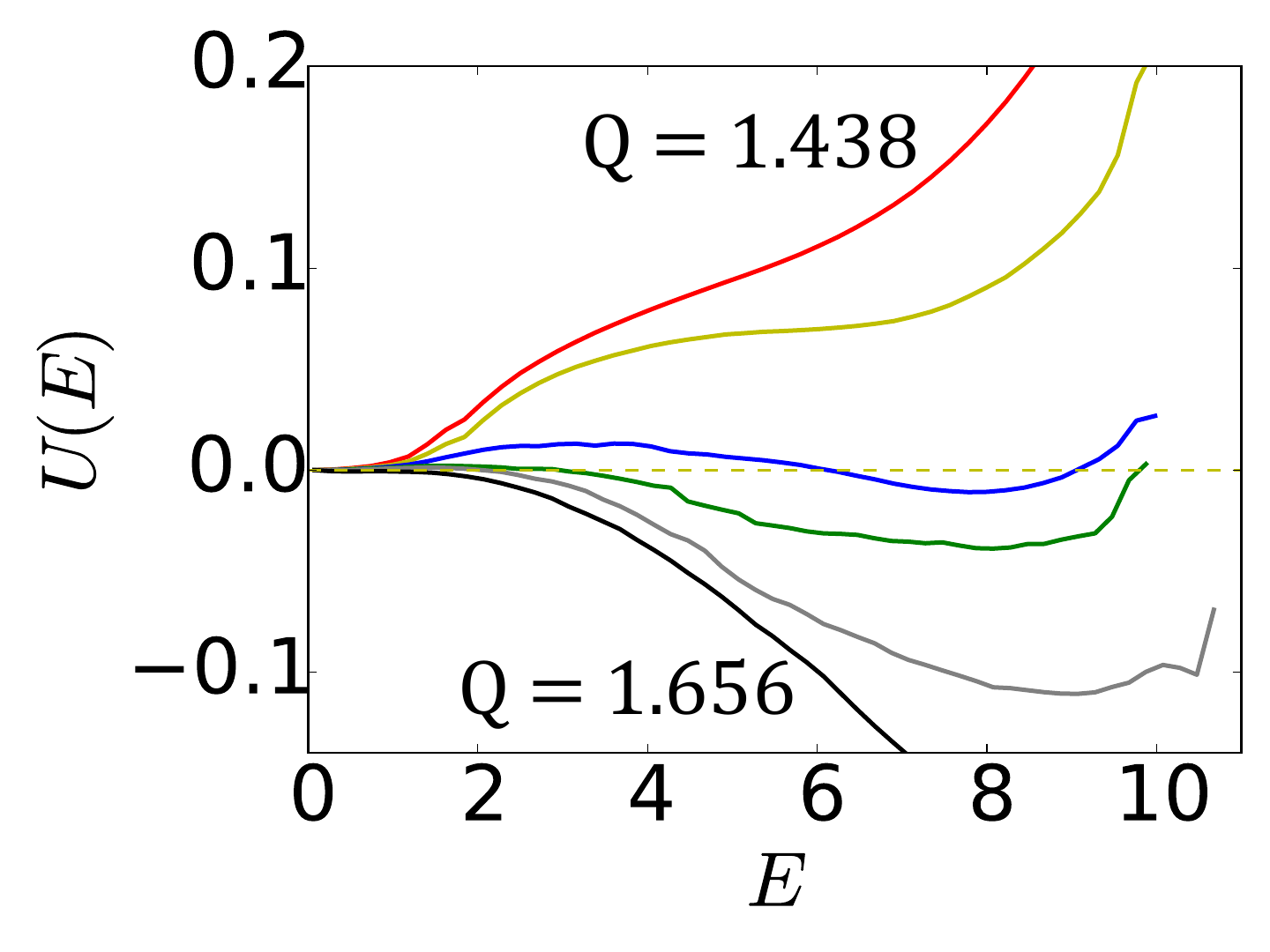} 
\end{subfigure}
\begin{subfigure}[hb]{0.32\linewidth}
\includegraphics[width=1\textwidth]{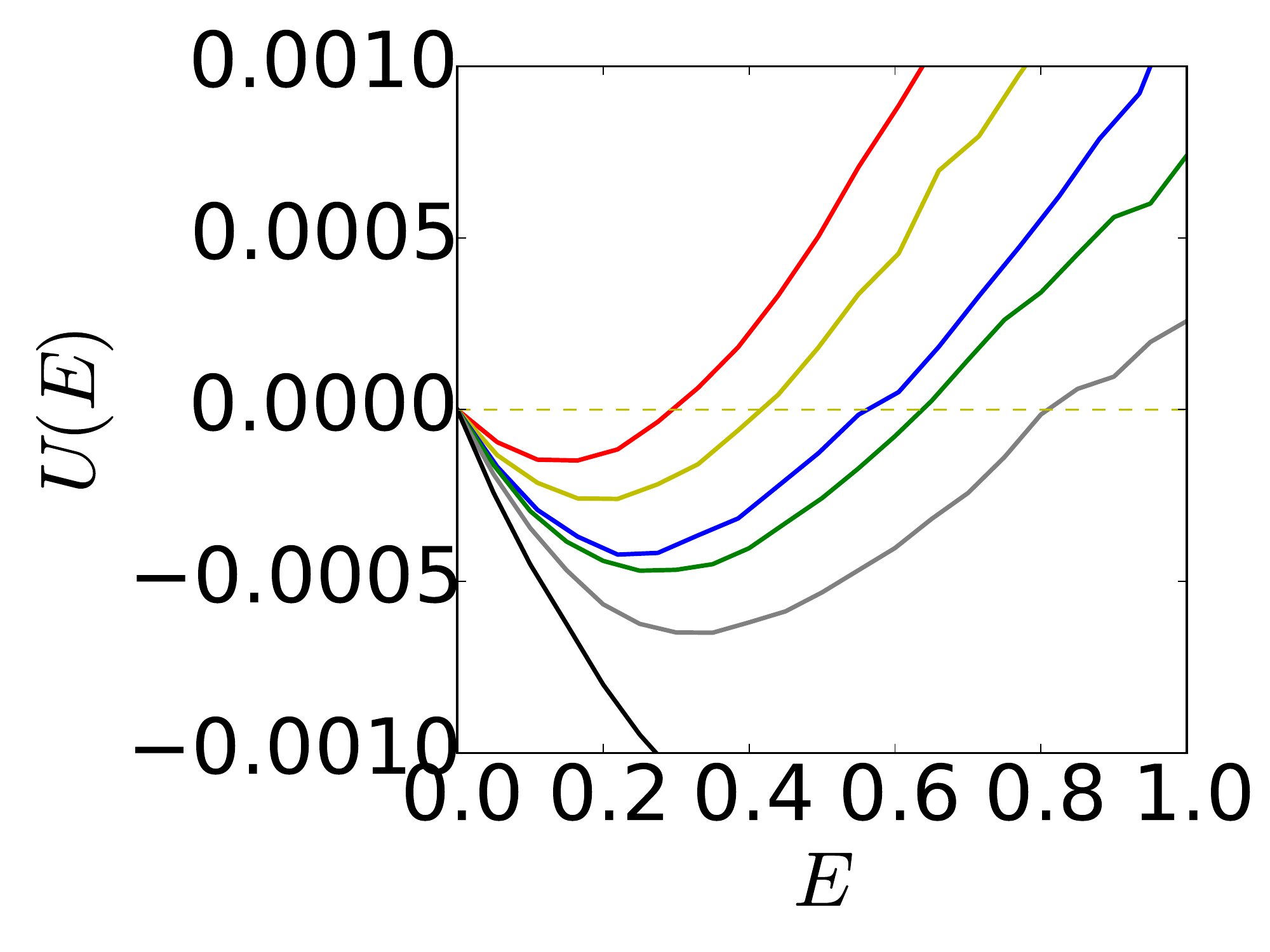} 
\end{subfigure}
\begin{subfigure}[hb]{0.32\linewidth}
\includegraphics[width=1\textwidth]{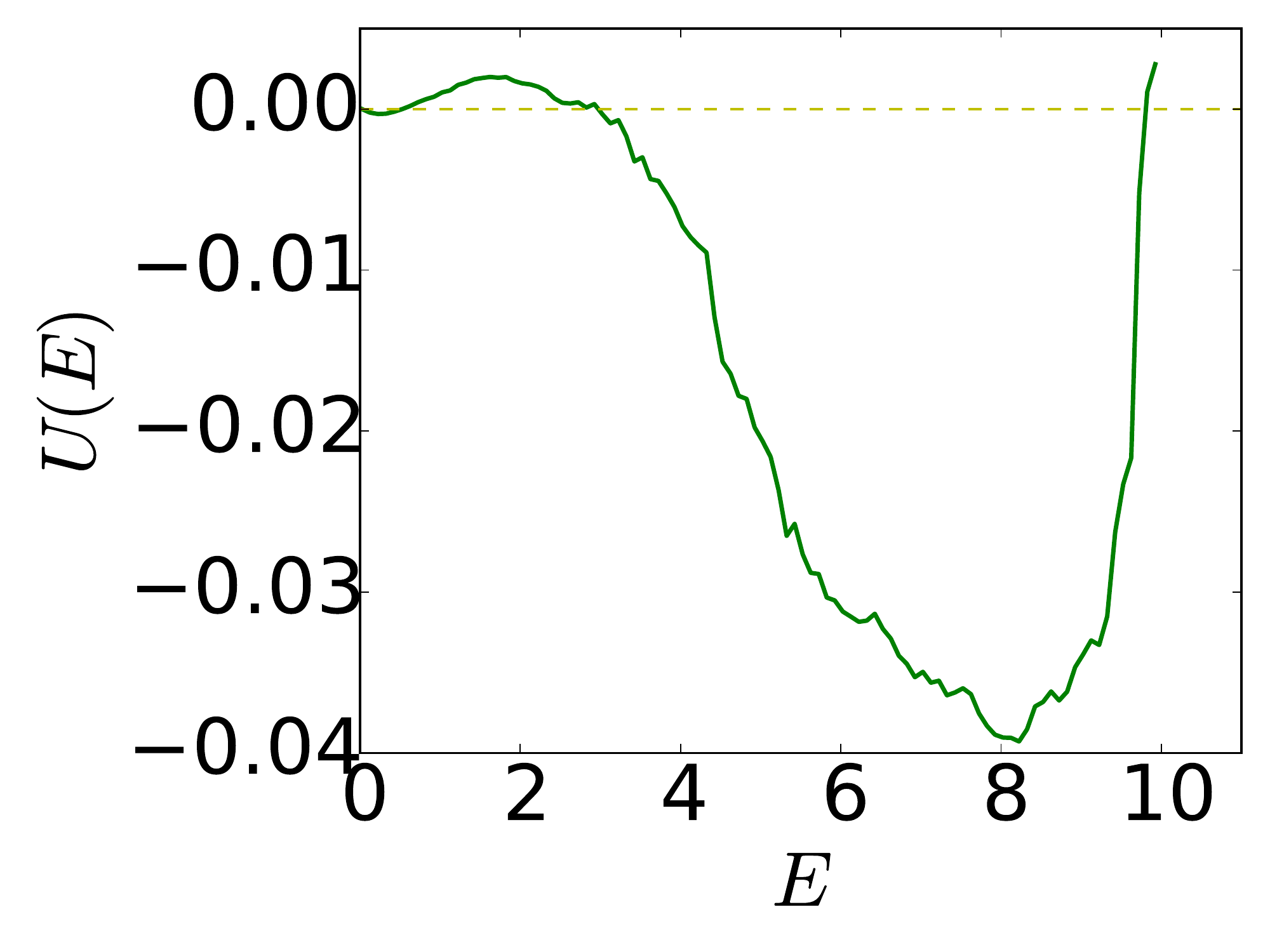} 
\end{subfigure}
\caption{ Left panel: The potential $U(E)$ obtained from DNS (both decay and build-up experiments) for different values of $Q$. The values of $Q$ from top to bottom are $1.438$ (red), $1.5$ (yellow), $1.556$ (blue), $1.563$ (green), $1.594$ (grey), $1.656$ (black).
Middle panel: an enlarged view close to $E=0$. 
Right panel: an enlarged view of $U(E)$ for $Q=1.563$.
The potential is obtained using eq. \ref{eq:Ameasure} for stochastic forcing. 
} 
\label{fig:A_measured}
\end{figure}

\begin{figure}
\centering
\begin{subfigure}[hb]{0.35\linewidth}
\includegraphics[width=1\textwidth]{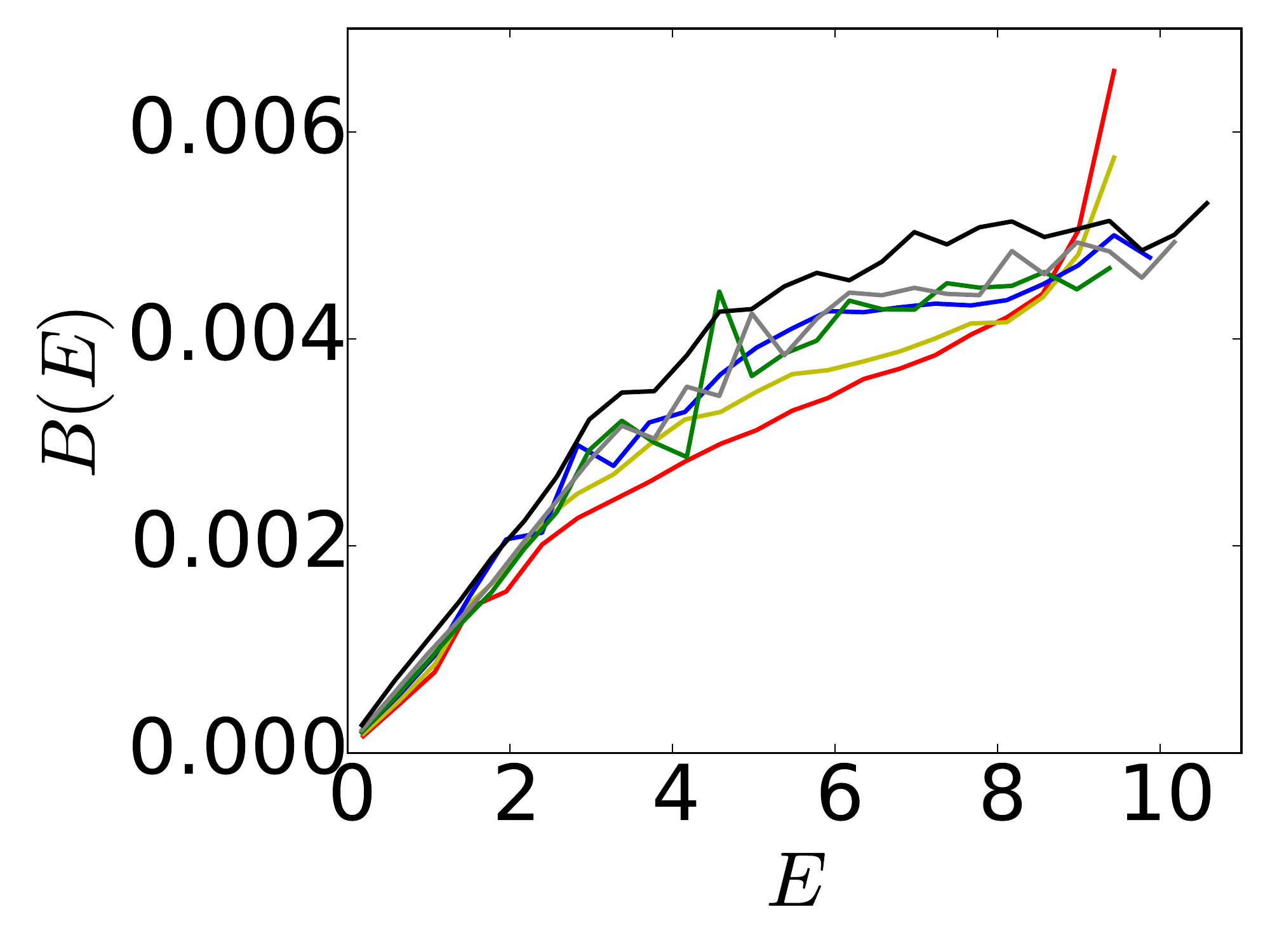} 
\end{subfigure}
\begin{subfigure}[hb]{0.35\linewidth}
\includegraphics[width=1\textwidth]{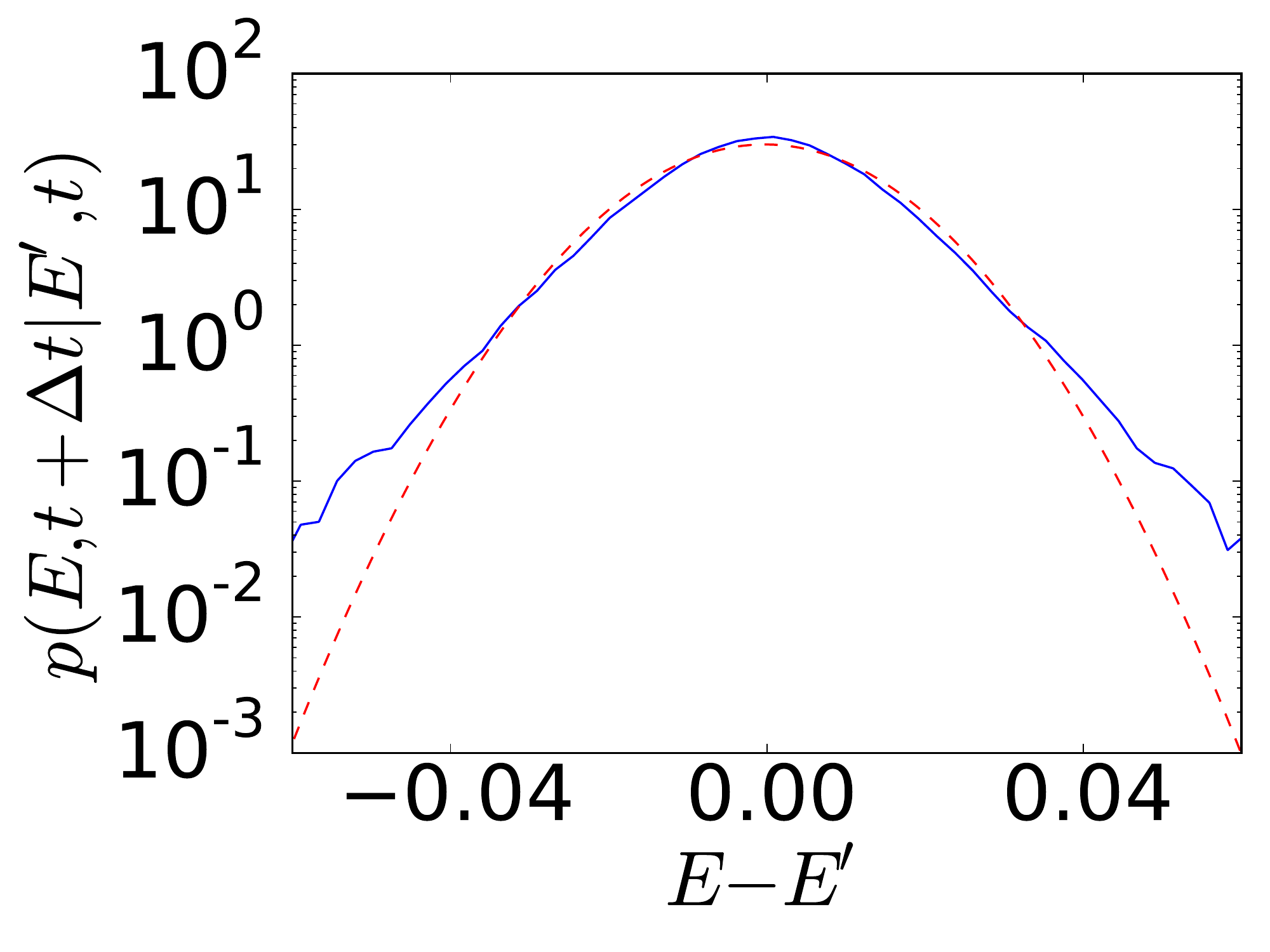} 
\end{subfigure}
\caption{ Left panel: The $E$-dependent diffusion constant $B(E)$ obtained 
using eq. \ref{eq:Bmeasure} for stochastic forcing. 
Colors are as in Fig~\ref{fig:A_measured}. 
Right panel: Transition probability $p(E,t+\Delta t|E^{\prime},t)$ obtained from DNS (blue solid line) and a Gaussian fit
(red dashed line). We set
$E'=0.5$ and $\Delta t=0.2$ with $Q=1.563$ for a stochastic forcing build-up experiment. 
} 
\label{fig:B_measured}
\end{figure}

Motivated by this observation, in this section, we discuss to what extent the dynamics of $E\equiv V_{ls}^2/(\epsilon \ell)^{2/3}$ can be described by one-dimensional Markov process.  
Assuming the continuity of the trajectory of $E(t)$, a general form of this process is written as \tn{the following Langevin equation} \citep{gardiner1986handbook}: \tn{denoting by $\Delta E(t)$ the increment of $E(t)$ during a small time interval $\Delta t$, the equation is written as}
\begin{equation}
\Delta E(t) = - \frac{\partial U(E)}{\partial E} \Delta t + \sqrt{2 B(E)} \Delta W(t),
\label{eq:E_2Langevin}
\end{equation}
\tn{where $\Delta W(t)$ is a Gaussian noise satisfying $\langle \Delta W(t)\rangle = 0$ and $\langle \Delta W(t) \Delta W(t^{\prime})\rangle = \Delta t \ \delta_{t,t^{\prime}}$ and we use the It\^o rule for the multiplication between $\Delta W(t)$ and $\sqrt{2B(E)}$, {\it i.e.}, $\langle  \sqrt{2 B(E)} \Delta W(t) \rangle  = 0$.}
The potential $U(E)$ and the $E$-dependent diffusion constant $B(E)$ characterise the dynamics and are to be determined. Note that the noise term  models the smaller-scale turbulent motions. 
Eq.~(\ref{eq:E_2Langevin}) indicates that the average and the variance of 
\tn{$\Delta E(t)$}
conditional on $E$ are equal to 
\tn{$- \left ( \partial U(E) / \partial E \right ) \Delta t$ and $2 B(E) \Delta t$} 
respectively. From the time-series data of the DNS, we thus evaluate these  statistical properties and estimate $U(E)$ and $B(E)$ (see Appendix \ref{appA} for more detail). The results are shown in figures \ref{fig:A_measured} and \ref{fig:B_measured}. When $Q$ is well below the transition, $U(E)$ has a single minimum at $E\simeq0.1$. Positive slopes at larger values of $E$ indicate a mean drift towards the \tn{3-D flow state}. As $Q$ approaches the transition, $U(E)$ becomes flatter, meaning that this drift vanishes, and a second minimum appears at $E\simeq 7$ for $Q\simeq 1.5$. This indicates that the system is in a bistable regime. The second minimum becomes more pronounced as $Q$ is further increased. Eventually for $Q \gtrsim 1.65$ the first minimum disappears and the system is left with a single minimum corresponding to the condensate state. 
\tn{The time scale for the system to jump from one minimum to the other can be estimated using a first-passage time formula in terms of $U(E)$ and $B(E)$ [see Section 5.2.7 ``First Passage Times for Homogeneous Processes'' in \citep{gardiner1986handbook}]. For example, we consider a transition time for build-up events at $Q=1.563$, where the corresponding $U(x)$ and $B(x)$ are shown in the right panel of figure~\ref{fig:A_measured} and in the left panel of figure~\ref{fig:B_measured}, respectively. The build-up event is then characterized as the jump from the potential (local) minimum around $E_{\rm min}=0.3$ to the potential maximum around $E_{\rm max}=1.8$. In this range of $E$, $B(E)$ can be well-described as a linear function $B(E)=bE$ (with $b\simeq 0.001$). We also approximate $U(E)$ as a linear function $U(E)=aE+c$ (with $a\simeq 0.002$) for simplicity\footnote{\tn{A generalization for more realistic potentials is straightforward \citep{gardiner1986handbook}. Here we choose this approximation because the obtained formula (\ref{eq:firstpassage}) becomes simple.}}. Using the first-passage time formula\footnote{\tn{We set a reflecting boundary at $E=E_{\rm min}$ and an absorbing boundary at $E=E_{\rm max}$. Substituting the explicit expressions of $B(E)$ and $U(E)$ into the mean first passage time formula in p.139 of \citep{gardiner1986handbook}, we get (\ref{eq:firstpassage}). Note that, for more detailed analysis, we need to model $B(E)$ and $U(E)$ outside $E_{\rm min}<E<E_{\rm max}$, and define a reflecting boundary at $E=0$ and an absorbing boundary at another local minimum of $U(E)$ corresponding to the 2-D stable state. This will possibly increase the estimation of $\tau_b$ compared with (\ref{eq:firstpassage}).} }, we get the time scale $\tau_b$ as
\begin{equation}
\tau_b = \frac{2}{ a + b}\left [   E_{\rm min} - E_{\rm max} + 
  \frac{b}{a} E_{\rm max} \left ( -1 + (E_{\rm max}/E_{\rm min})^{a/b} \right ) \right ],
\label{eq:firstpassage}
\end{equation}
which estimates $\tau_b = 2\times 10^4$. This shows a rough agreement with the data
in the left panel of figure \ref{fig:mean_time} in the previous section.}

The interpretation of our results using a Langevin equation with a double-well potential is the simplest model to describe a discontinuous phase transition. The presence of a double-well potential implies that the system can always jump from one state to the other if one waits long enough. However 
we cannot conclude that 
$\langle \tau_d \rangle$ and $\langle \tau_b \rangle$ do not diverge for finite $Q$ (Scenario 4 in figure \ref{fig:scenarios}) from this observation.
First of all, the range of $Q$ that we have studied is highly limited. Second, 
an oversimplification has been made when we assumed that the system could be described by a single variable $E$, while the system in reality evolves in an extremely high-dimensional space. 
Indeed, disagreements between the 1D model (\ref{eq:E_2Langevin})
and DNS results can be detected when we look at the distribution of the energy increment $\Delta E$ over a small time interval $\Delta t$. According to the Langevin equation (\ref{eq:E_2Langevin}), this distribution needs to be Gaussian, but this is not exactly true for the DNS results: using the time-series data of DNS, we evaluate the transition probability $p(E,t+\Delta t|E^{\prime},t)$ from the state $E^{\prime}$ at time $t$ to $E$ at time $t+\Delta t$.
In figure \ref{fig:B_measured} we plot this probability with $E'=0.5$ and $\Delta t=0.2$ for stochastic forcing with $Q=1.5625$, showing deviations from a Gaussian distribution. 


\section{Conclusions}
\label{sec:conclusions}

In this work we have studied the statistical properties of thin-layer flows close to the transition between a \tn{3-D flow state} and the formation of a 2-D condensate. Such transitions have recently been discovered in a variety of systems \citep{seshasayanan2018condensates,favier2019subcritical,vankan2018condensates} and 
this work is the first attempt to systematically study their statistics.
We have measured the probabilities of the transition times between the two states, 
where the mean 
transition times
were shown to increase by three orders of magnitude by a relatively small change (10\%) of the control parameter $Q$.
We 
point out the qualitative similarity between figure 5 of \citep{avila2011onset} for the turbulent-laminar transitions in pipe flows and figure \ref{fig:mean_time} of our work. Although the physical situations are different, in both cases we observe super-exponentially growing time scales of two competing processes. Our results could neither exclude nor confirm whether this sharp increase has a double-exponential scaling form $\exp(\beta\exp(\alpha x))$,  
supported by an argument based on extreme value statistics \citep{goldenfeld2010extreme}.
This leaves a possibility of the divergence of mean transition times, {\it i.e.}, transitions from one state to the other could become impossible for a certain range of $Q$.

Our results show that the system can be modeled to some extent as an inertia-less particle trapped in a one-dimensional potential in the presence of stochastic noise. The model revealed that close to the transition the potential displays two minima implying the existence of a bistable state.
Discrepancies in the noise statistics of the DNS and of the stochastic model were observed, that were attributed to the multi-dimensional nature of the real problem.

Several simplifying assumptions have been made in this work in order to render the problem more tractable.
For example, the domain was triply periodic and also the forcing was 2-D. These simplifications could limit the applicability of these results to laboratory or natural flows. Further investigations with more realistic boundary conditions and forcing 
are thus necessary.
More importantly, the present investigation was limited to a single, relatively small, value of
$\Rey$ and $K$ (the scale separation between the forcing length scale and the horizontal domain
scale).
Examining the observed behaviour at larger values of $\Rey$ and $K$ is crucial for validating the 
robustness of our results and for their applications to natural flows. \tn{Another important limitation is the low vertical resolution of only 16 points imposed by the long simulation times. It is known that the minimum resolution required for correct DNS of turbulence is around 256, for a detailed study of resolution issues in turbulence see \citep{donzis2008dissipation}. The impact of the low vertical resolution used here is that the small-scale vertical motion will be more strongly affected by viscosity. This likely affects the statistics of the rare transitions, the latter being induced by turbulent fluctuations. We use the term ''3-D flow state'' instead of ''3-D turbulent state'' to highlight this.} Unfortunately, the rareness of decay and build-up events close to the bistable regime limits the range of parameters and the resolution that can be examined with DNS.

\tn{However in recent years, alternative methods have been developed that could address this problem.}
In particular, \tn{our understanding of the problem} could benefit from studies using rare event sampling algorithms, such as a method calculating the instanton based on Freidlin-Wentzell theory \citep{Chernykh2001, Heymann2008, Grafke_2015_2, Grafke_2015, Grigorio_2017}, splitting methods that copy rare event realizations to efficiently accumulate statistics \citep{Allen_2005, Giadina_2006, cerou2007adaptive, tailleur2007probing, Teo_2016, Nemoto_2016, Lestang_2018, bouchet2018rare} and also a recently proposed method that relies on feedback control of Reynolds number \citep{PhysRevE.97.022207}. 
Such studies can help to overcome the difficulty caused by the extremely long computational time required to accurately describe the rare transition events close to the onset. Studies at larger $\Rey$ or scale separations could therefore become tractable using these methods. \tn{In addition to more efficient simulation, statistics at higher Reynolds numbers may be accessible if one chooses not to study transitions but bursting events out of one of the attractors. Bursts are more frequent than transitions in the bistable regime and their statistics therefore are more easily accessible.}

Furthermore, given the large experimental literature \citep[see][]{xia2011upscale,xia2017two} on the transition between 3-D turbulence and condensation in thin layers, it would be exciting and very important to study the observed bistability experimentally. The biggest advantage of an experiment compared to DNS would be that much longer observation times are possible as well as higher $\Rey$. The same remarks apply to rotating turbulence.

\section*{Acknowledgements}
\tn{The authors gratefully acknowledge the comments of three anonymous referees, which have helped improve the clarity of the paper.} This work was granted access to the HPC resources of MesoPSL financed by the R\'egion Ile de France and the project Equip@Meso (reference ANR-10-EQPX-29-01) of the programme Investissements d’Avenir supervised by Agence Nationale pour la Recherche and HPC resources of GENCI-TGCC-IRENE \& CINES Occigen (Project A0030506421 \& A0050506421) where the present numerical simulations have been performed. This work has also been supported by the Agence  nationale de la recherche (ANR DYSTURB project No. ANR-17-CE30-0004). We thank Freddy Bouchet for helpful discussions.
\appendix
\section{Determining $U(E)$ and $B(E)$}\label{appA}
In this appendix we express $U(E)$ and $B(E)$ in terms of transition probabilities to obtain expressions which allow determining them from the DNS time series. For convenience, let $A(E) \equiv -\frac{\mathrm{d} U}{\mathrm{d} E}$, the drift velocity. For a small time increment $\Delta t$, we denote by $p(E,t+\Delta t|E^{\prime},t)$ the transition probability \tn{from $E(t)=E^{\prime}$ at time $t$ to $E(t+\Delta t)=E$ at time $t+\Delta t$}. It obeys, \citep[see][]{gardiner1986handbook},
\begin{equation}
p(E,t+\Delta t|E^{\prime},t) - \delta (E - E^{\prime})
 = - \frac{\partial}{\partial E}A(E) \delta (E - E^{\prime}) \Delta t  + \frac{\partial^2}{\partial E^2} B(E)\delta (E - E^{\prime}) \Delta t + O(\Delta t^2)
\end{equation}
By multiplying both sides by $E-E^{\prime}$ or  $\left (E-E^{\prime}\right )^2$ and integrating over $E$, we get
\begin{eqnarray}
\int dE (E - E^{\prime}) p(E,t+\Delta t|E^{\prime},t)
 = & A(E^{\prime}) \Delta t + O(\Delta t^2), \label{eq:Ameasure} \\
\int dE \left (E-E^{\prime}\right )^2 p(E,t+dt|E^{\prime},t)
 = & 2 B(E^{\prime}) \Delta t + O(\Delta t^2). 
 \label{eq:Bmeasure}
\end{eqnarray}
The left-hand sides (and thus $A$ and $B$) are measurable from a time-series $E_{DNS}(t)$ of large-scale energy 
by computing the mean of $E_{DNS}(t+\Delta t)-E_{DNS}(t)$ and $(E_{DNS}(t+\Delta t)-E_{DNS}(t))^2$, over all times $t$ for which $E_{DNS}(t)=E^\prime$, for all values of $E^\prime$. Finally the potential $U(E)$ is obtained from $A(E)$ by simple integration.

\bibliographystyle{jfm}
\bibliography{condensate_statistics}

\end{document}